\documentclass[12pt,a4paper]{article}   
\usepackage[dvips]{graphicx}
\usepackage{amsmath}
\usepackage{amssymb}
\usepackage{amsthm}
\usepackage{amscd}

\def\m@th{\mathsurround=0pt}

\def\fsquare(#1,#2){
\hbox{\vrule$\hskip-0.4pt\vcenter to #1{\normalbaselines\m@th
\hrule\vfil\hbox to #1{\hfill$\scriptstyle #2$\hfill}\vfil\hrule}$\hskip-0.4pt
\vrule}}

\newcommand{\np}{\nonumber\\}
\newtheorem{thm}{Theorem}

\title{
The dilute $A_L$ models and the integrable perturbations of unitary minimal CFTs. 
}
\author{ J. Suzuki\thanks{e-mail: sjsuzuk@ipc.shizuoka.ac.jp}\\
        \parbox{0.9\textwidth}{
        {\em
        \begin{center}
      Department of Physics,\\
      Faculty of Science, Shizuoka University \\
      Ohya 836, Shizuoka\\
       Japan
        \end{center}
        }}
       }
\date{May  2003}
\begin{document}
\maketitle
\begin{abstract}
Recently, 
a set of thermodynamic Bethe ansatz equations is proposed by Dorey, Pocklington and Tateo  for 
 unitary minimal models perturbed by $\phi_{1,2}$ or $
 \phi_{2,1}$ operator.
We examine their results in view of
the lattice analogues, dilute $A_L$ models at regime 1 and 2.
Taking  $M_{5,6}+\phi_{1,2}$ and $M_{3,4}+\phi_{2,1}$ as the
simplest examples, we will explicitly show that the conjectured TBA equations can be
recovered from the lattice model in a scaling limit.
\end{abstract}

\clearpage

\section{Introduction}

Since the breakthrough in 
the integrable perturbation theory of CFT  \cite{Zam891, Zam892}, 
there has been a lot of progress in the understanding of  $\phi_{1,3}$ perturbation
theory \cite{EguchiYang, AlZamoRSOS}.
On the other hand, 
although the  remarkable example, the Ising model in a magnetic field , 
was treated in \cite{Zam891},  the progress on 
 the $\phi_{1,2}$ and $\phi_{2,1}$  perturbed theories has been steady but slow.

The systematic studies on the bootstrap procedure on $S$ matrix have been
initiated in \cite{Smir} and \cite{ChimZam}. 
The latter approach, based on the scaling $q-$ state Potts field theory, has been
further elaborated by Dorey et al \cite{DTP1}.
Thanks to the Coleman-Thun mechanism,  they argue that  the contributions from spurious poles
cancel and conclude  the closed  set of $S-$ matrices for a wide range of parameters.

The check of the results against a finite size system, however suffers from the non-diagonal nature
of the scattering process.   
Due to the lack of a relevant string hypothesis,  the diagonalization of the transfer matrix is far from trivial.
In \cite{DTP2},  conjectured is  a set of thermodynamic Bethe ansatz equations  (TBA)  from the 
consideration on special cases of which they found similarity to the TBA for the sine-Gordon model .
Roughly speaking, they proposed the TBA by gluing the ``breather-kink" part and the ``magnon" part
in which the latter originates from the sine-Gordon  model at specific coupling\cite{Roberto95, DTT}.
Although the derivation is intuitive, the resultant equations pass many non trivial checks.

In this report, we shall examine the problem
in view of a solvable lattice model .
As a lattice analogue to  $M_{L, L+1}+ \phi_{1,2},  M_{L+1, L+2}+ \phi_{2,1}$ 
we consider   the $L-$ state RSOS model proposed  in \cite{WNS1, WNS2},
which will be referred to as the dilute $A_L$ model.
There are several evidences for this correspondence, the central charge \cite{WNS1}, the scaling dimensions
of the leading perturbation\cite{WNS1, WPNS},  universal ratios \cite{BS1, KSeaton, KorffSeaton} and so on.

The question whether it shares the identical TBA to
describe its finite temperature (size) property has not  yet been fully
answered.
The purpose of this report is to  present   positive evidences for this inquiry.

There are already  few examples to the demonstration of the  equivalence.
The common TBA of  the dilute $A_3$ model  at regime 2 and
the $M_{3,4}+\phi_{1,2} $ case is firstly 
proved in \cite{BWN}.
There the most dominant solutions to the Bethe ansatz equation are explicitly
identified in the form of the ``string solution" , which
 leads to the famous $E_8$  TBA.
In the case $L=4,6 $, corresponding to the $E_7, E_6$ case,  
 such explicit identification of  string hypothesis 
 seems not yet to  be completed.

An alternative approach,  based on the quantum
transfer matrix (QTM) \cite{Suz85,Klu92},  has been  successfully applied to 
$L=3,4,6 $\cite{JSE8, JSE7}.
The functional relations among properly chosen QTMs play the fundamental role there
and it enables to derive TBA without knowing the explicit locations of
dominant solutions to Bethe ansatz equation .

For $L=3,4,6 $ cases, the underlying  affine Lie algebraic structure 
($E_8, E_7, E_6$, respectively)
provides
several clues in the investigation of the functional relations among QTMs.
The remaining case, which seems to lose a direct connection to affine Lie algebra
in general (see, however exceptions \cite{DTP2}).
It might be thus challenging to the clarify the functional relation, and thereby
see if the $Y$ system in is actually recovered.
In this report, 
 the last ``exceptional" case (in terminology of  \cite{DTP2})  $M_{5,6}$ for the $\phi_{1,2}$ perturbation, 
and the first  exceptional case  $M_{3,4}$ for the $\phi_{2,1}$ perturbation are focused.

This paper is organized as follows.
In the next section, we give a brief review on the dilute
$A_L$ models and the QTM method.
Section 3 is devoted to the discussion on  the dilute $A_5$ model
at regime 2 which is expected to be a lattice analogue of 
the $M_{5,6}+\phi_{1,2}$ theory.
Fusion QTMs parameterized
by skew Young diagrams are introduced and found to 
satisfy a set of closed functional relations.
 It will be shown that the conjectured TBA 
 is naturally derived in a scaling limit.
 In case of  the dilute $A_L$ model, $L$ even,  a fundamental role seems 
 to be played by a  ``kink" transfer matrix. As the simplest and the most well-known
 example, we treat   $M_{3,4}+\phi_{2,1}$,  corresponding to
 the Ising model off critical temperature, in section 4.
We conclude the paper with brief summary and discussion in section 5.

%
%
\section{The dilute $A_L $ model and the quantum transfer matrix}

\begin{figure}[hbtp]
\centering
\includegraphics[width=6cm]{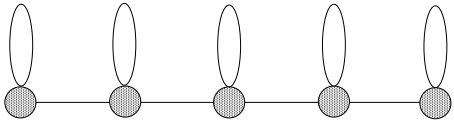}
\caption{ 
An incidence diagram for the dilute $A_5$ model. The local states corresponding to
connected nodes can be located to nearest neighbor sites on a square lattice.
  }
\label {adj}
\end{figure}

The dilute $A_L$ model is proposed in \cite{WNS1} as
an elliptic extension of the Izergin-Korepin model \cite{IK}.
The model is of the restricted SOS type with local 
variables $\in \{1,2,\cdots,L \}$.
The variables $\{a, b\} $ on neighboring sites
should satisfy the adjacency condition, $|a-b|\le 1$,
which is often described by a graph in fig.\ref{adj}.
%
%
%
%
In \cite{WNS1}, the RSOS weights,  satisfying the Yang-Baxter relation,  have
been found to be parameterized by the spectral parameter $u$ and the elliptic nome $q$.
The crossing parameter $\lambda$ needs to be a function of $L$ for the restriction. 
The model exhibits four different physical regimes depending on parameters,
\begin{itemize}
 \item regime 1. $0<u<3 , \,\lambda=\frac{\pi L}{4(L+1)},\, L\ge 2 $
 \item regime 2. $0<u<3, \, \lambda=\frac{\pi (L+2)}{4(L+1)},\,  L \ge 3$
 \item regime 3. $3-\frac{\pi}{\lambda}<u<0, \,
  \lambda=\frac{\pi (L+2)}{4(L+1)}, \, L \ge 3 $
 \item regime 4.  $3-\frac{\pi}{\lambda}<u<0,\, 
  \lambda=\frac{\pi L}{4(L+1)}, \, L \ge 2$.
\end{itemize}
We are interested in  regimes 1 and 2.

The central charge and scaling dimension associated to
leading perturbation evaluated in \cite{WNS1, WPNS}
suggests,
\begin{itemize}
 \item The dilute  $A_{L-1} $ model at regime 1 is an underlying lattice theory for 
            $M_{L, L+1}+\phi_{2,1}$ 
\item The dilute  $A_{L} $ model at regime 2 is an underlying lattice theory for 
            $M_{L, L+1}+\phi_{1,2}$ 
\end{itemize}
There are also further evidences supporting this correspondence, as mentioned in the introduction.

One can introduce an associated  
 1D quantum system to the above 2D classical model.
 The Hamiltonian ${\cal H}_{\rm 1D}$ for the former is defined from the row to row transfer matrix
 $T_{\rm RTR}(u)$ 
 of the latter,   by
 
\begin{equation*}
{\cal H}_{1D} = \epsilon \frac{\partial}{\partial u} \log  T_{RTR}(u) |_{u=0}.
\end{equation*}

We omit the explicit operator form  of ${\cal H}_{1D}$.
The parameter $\epsilon=-1,(1)$ labels regimes 1 and 2 (3 and 4).

The thermodynamics of the  1D quantum system is the central issue in the following.
We apply the method of QTM \cite{Suz85, Klu92} to this problem.
Leaving details to references, we list the only relevant  results for the following  discussion.

A fundamental  QTM is defined in a staggered manner 
$$
(T_{\hbox{\footnotesize{QTM}}}(u,x))^{\{b\}}_{\{a\}} =\prod_{j=1}^{N/2}
\mbox{\parbox[c][0.9cm]{0cm}{}}^{b_{2j-1}}_{a_{2j-1}}
\fbox{\parbox[c][0.7cm]{0.7cm}{ {\scriptsize $u\!+\! ix$}  }} \>
\mbox{\parbox[c][0.9cm]{0cm}{}}^{b_{2j}}_{a_{2j}}
\> \>
\mbox{\parbox[c][0.9cm]{0cm}{}}^{\,\,a_{2j+1}}_{a_{2j}}
\fbox{\parbox[c][0.7cm]{0.7cm}{ {\scriptsize $u\!-\! ix$}  }} \>
\mbox{\parbox[c][0.9cm]{0cm}{}}^{b_{2j+1}}_{b_{2j}}.
$$
In the above, squares represent  Boltzmann weights; four indices
represent local variables and the spectral parameters are 
specified inside of them.
The fictitious dimension $N$( even) , 
 sometimes referred to as the Trotter number,  is introduced.
 It has nothing to do with  the real system size of the original 1D system.
 The real system size will not appear in our discussion as the quantities
after taking the thermodynamic limit is of our interest.

It is vital that  two (spectral) parameters $u, x$ exist and that only the latter  
concerns  the commutative property of QTMs,
$
[T_{\hbox{\footnotesize{QTM}}}(u,x), T_{\hbox{\footnotesize{QTM}}}(u,x') ]=0.
$
The remaining  parameter $u$ plays the role in intertwining the finite Trotter number ($N$) system 
and the finite temperature system  ($\beta$)  by  $u= u^{\ast} =-\epsilon \frac{\beta}{N}$.
More concretely, the free energy per site is represented {\it only} by the
largest eigenvalue  $T_1(u,x)$  of $T_{\hbox{\footnotesize{QTM}}}$ at
$x=0$ and    $u= u^{\ast}$,
\begin{equation*}
\beta f = -\lim_{N\rightarrow \infty}\log  
 T_1 (u^{\ast}, x=0) \Bigr .
\end{equation*}
The eigenvalue $T_1(u,x)$ 
takes the form
\begin{eqnarray}
T_1(u,x)&=& 
w \phi(x+\frac{3}{2}i) \phi(x+\frac{1}{2}i)\frac{Q(x-\frac{5}{2} i)}{Q(x-\frac{1}{2} i)} +
\phi(x+\frac{3}{2}i) \phi(x-\frac{3}{2}i) 
\frac{Q(x-\frac{3}{2}i)\, Q(x+\frac{3}{2} i)}{Q(x-\frac{1}{2} i)\, Q(x+\frac{1}{2} i)} \np
&+ &w^{-1} \phi(x-\frac{3}{2}i) \phi(x-\frac{1}{2}i)  
 \frac{Q(x+\frac{5}{2}i)}{Q(x+\frac{1}{2} i)},  \label{dvf}  \\
Q(x)&:=&\prod_{j=1}^{N} h[x-x_j] \np
\phi(x) &:=& \Bigl(\frac{h[x+(\frac{3}{2}-u)i] h[x-(\frac{3}{2}-u)i]}{h[2i] h[3i]}\Bigr)^{N/2},
\qquad h[x] := \vartheta_1(i\lambda x),  \nonumber
\end{eqnarray}
where  $w=\exp(i \frac{ \pi \ell}{L+1} )   $ ($\ell=1$ for the largest eigenvalue
sector).

The parameters, $\{ x_j \}$ are solutions to ``Bethe ansatz equation" (BAE),
\begin{equation}
w \frac{\phi(x_j+i)}{ \phi(x_j-i)}=
-\frac{Q(x_j-i) Q(x_j+2 i)}{Q(x_j+i)Q(x_j-2 i)},
\quad j=1,\cdots, N.
\label{bae}
\end{equation}
From now on we suppress the dependency on $u$ which must be
 set as $u=u^{\ast}$.

It has been shown in many examples \cite{Klu93}, that the functional relations among 
``generalized"  (fusion) QTMs offer a way to evaluate the free energy 
without precise knowledge on the locations $\{x_j \}$.
We adopt the same strategy here and  shall discuss the 
functional relations realized among  fusion QTMs of  the dilute $A_L$ model
below.
 
 %
 \section{QTM associated to skew Young diagrams and quantum Jacobi -Trudi
 formula}
 %
 
We introduce fusion QTMs associated to Young diagrams.
 The idea to connect  Young diagrams and (eigenvalues of)  QTM,
 originated in \cite{BR, SuzG2, KS} is very simple.
Let three boxes with
letters 1,2 and 3 represent the three terms in
eigenvalue of the quantum transfer matrix (\ref{dvf}), 
$$
T_1(x)= \framebox[0.4cm][c]{1}_{x} + \framebox[0.4cm][c]{2}_{x} +
\framebox[0.4cm][c]{3}_{x}. 
$$
Obviously, the eigenvalue of a fusion QTM can be represented by
a summation of  products of ``boxes" with different letters and spectral parameters,
over a certain set.
The point is that 
the set can be identified with semi-standard 
Young tableaux (SST) for $sl_3$.
 We state the above situation more precisely.
Let $\mu$ and $\lambda$ be a pair of Young tableaux satisfying 
 $\mu_i \ge \lambda_i, \forall i$.
We subtract  a diagram  $\lambda$ from $\mu$, which is called 
a skew Young diagram $\mu-\lambda$.
The usual Young diagram is the special case that $\lambda $ is empty, and
we will omit $\lambda$ in the case hereafter.
On each diagrams, the spectral parameter changes $+2i$ from the left  box to the right and
$-2i$ from the top box to the bottom. 
We fix the spectral parameter associated to  the right-top box to be $x+i(\mu'_1+\mu_1-2)$
(or equivalently   the spectral parameter associated to  the left-bottom  box to be $x-i(\mu'_1+\mu_1-2)$ ).
Insert a letter $\ell_{i,j}$ to the $(i,j)$ -th box such that the  semi-standard  condition is satisfied.
We denote its spectral parameter by $x_{i,j}$.
Then the product  
$$
\prod_{i,j}  \framebox[0.6cm][c]{$\ell_{i,j}$}_{x_{i,j}}
$$
is associated to the Young table.  
The summation over the tableaux satisfying the  semi-standard  condition then defines

\begin{equation}
{\cal T}^{\vee}_{\mu/\lambda}(x)=\sum_{ \{\ell_{i,j} \} \in   {\rm SST} } 
 \prod_{i,j}  \framebox[0.6cm][c]{$\ell_{i,j}$}_{x_{i,j}}
\label{sumSST}
\end{equation}

which is expected to be the eigenvalue of a fusion QTM.

The simplest subset of the above is the QTM based on 
Young diagrams of  the rectangular shape. 
It was shown \cite{JSE8} that for any such member reduces to 
QTM  of $1\times m$  Young diagram,
which is related to $m-$ fold symmetric fusion.
For later convenience, we introduce a renormalized $1\times m$  fusion QTMs 
$T_m(x)$ by 
$$
T_m(x)= \frac{1}{f_m(x)}
\sum _{i_1 \le i_2 \le \cdots \le i_m}
\begin{tabular}{|l|l|l|l|}
\hline
$i_1$& $i_2$& $\cdots$& $i_m$ \\
\hline
\end{tabular}.
$$
The renormalization factor $f_m$, common to tableaux of width $m$, is given by
$$
f_m(x):= \prod_{j=1}^{m-1} 
\phi(x\pm i(\frac{2m-1}{2}-j)) .
$$ 
Hereafter,  for any function $f(x)$,  we denote by $f(x\pm ia)$ the product
$f(x+i a)f(x-ia)$ .

Then the  resultant $T_m$'s are all degree $2N$ w.r.t.  $h[x+{\rm shift }]$, and 
 have a periodicity due to Boltzmann weights;
$
T_m(x+P i) =T_m(x), 
$
where

\begin{equation}
P=
\begin{cases}
 \frac{4(L+1)}{L+2} ,& \hbox{ for regime 2} \\
 \frac{4(L+1)}{L}  ,&  \hbox{ for regime 1}.\\
\end{cases}
\label{period}
\end{equation}

Remarkably,  $T_m(x)$  enjoys a ``duality"
\begin{equation}
T_m(x)=
\begin{cases}
 T_{2L-1-m}(x), \,\, m=0,\cdots, 2L&, \hbox {for $L$ even } \\
T_{2L-1-m}(x+\frac{P}{2}i), \,\, m=0,\cdots, 2L&, \hbox {for $L$ odd }. \\
 \end{cases}
\label{dualonerow}
\end{equation}
This is deduced from the $a^{(2)}_2$ nature of the model and special choice of $\lambda$.
We have at least checked the validity numerically and assume their validity in this report.
The above two properties, the periodicity
 and the duality (\ref {dualonerow}) play the fundamental role in the 
proof of  the closed functional relations.

The real usefulness of  $T_m(x)$ lies in the fact that any 
QTM associated to a skew Young diagram can be represented in terms of 
their products.

\begin{thm}
Let ${\cal T}_{\mu/\lambda}(x)$ be a renormalized 
${\cal T}_{\mu/\lambda}(x)$ in  (\ref{sumSST})
by a common factor,
$
\prod_{j=1}^{\mu'_1} f_{\mu_j-\lambda_j} 
  (x+i(\mu_1' -\mu_1+\mu_j+\lambda_j-2j+1)).
$
Then the following  equality holds.
\begin{equation}
{\cal T}_{\mu/\lambda}(x) =
  {\rm det }_{1\le j,k\le \mu_1'} 
         ( T_{\mu_j-\lambda_k-j+k} 
            (x+i(\mu_1' -\mu_1+\mu_j+\lambda_k-j-k+1)) )
\label{qJT}
\end{equation}
where $T_{m<0}:=0$.
\end{thm}
We regard this as a quantum analogue of the Jacobi-Trudi formula.

By this, apparently ${\cal T}_{\mu/\lambda}(x)$  
 is an analytic function of $x$
due to BAE, and contains  the quantity of our interest,
$T_1(x)$ as a special case.
The former assertion is not obvious from 
the original definition by the tableaux, but
it is trivial from the quantum Jacobi-Trudi formula.

In the same spirit, we introduce  $\Lambda_{\mu/\lambda}(x)$, 
which is analytic under BAE,
 $$
 \Lambda_{\mu/\lambda}(x):=
 {\cal T}_{\mu/\lambda}(x) /.\{ T_{m\ge 2L}(x) \rightarrow 0 \}.
$$
The pole-free property of $\Lambda_{\mu/\lambda}(x)$ is apparent from 
(\ref{qJT}).

\section{ dilute $A_5$ model at regime 2 as a lattice analogue to $M_{5,6}+\phi_{1,2}$ }

For $M_{5,6}+\phi_{1,2}$, Dorey et al argued the existence of two kinds of particles,
2 kinks and 4 breathers.  For diagonalization of scattering theory, they introduced 2 magnons 
(massless particles),  in addition.
Explicitly, the $Y-$ system reads, 
\begin{eqnarray*}
Y_{B_1}(x\pm \frac{3}{14}i) &=& \Xi_{B_3}(x), 
\qquad
Y_{B_3}(x\pm \frac{3}{14}i) = \Xi_{B_1}(x) \Xi_{B_5}(x)
 \\
Y_{B_5}(x\pm \frac{3}{14}i) &=& \Xi_{B_3}(x) \Xi_{K_2}(x \pm  \frac{2}{14} i) \Xi_{K_1}( x) 
\Xi_1(x \pm \frac{1}{14}  i) \Xi_2(x) 
 \\
Y_{B_2}(x\pm \frac{3}{14}  i) &=&  \Xi_{K_1}(x \pm \frac{2}{14}  i) \Xi_1(x \pm \frac{1}{14}  i) \Xi_{K_2}(x) 
\\
Y_{K_2}(x\pm \frac{1}{14}  i) &=& \Xi_{B_5}(x)  {\cal L}^{(1)}(x),
\qquad 
Y_{K_1}(x\pm \frac{1}{14}  i) = \Xi_{B_2}(x) {\cal L}^{(1)}(x ) 
 \\
Y_1(x\pm \frac{1}{14}  i) &=& {\cal L}_2(x) {\cal L}_{K_2}(x){\cal L} _{K_1}(x), 
\qquad
Y_2(x\pm \frac{1}{14}  i) =  {\cal L}_1(x)  \\
\text{with} \qquad &\phantom{ }&  \\
{\cal L}_a(x) &:=& \frac{1}{1+ \frac{1}{Y_a  (x)  }} 
\qquad
 \Xi_a (x) :=1+Y_a(x)  
\end {eqnarray*}
where $a$ takes one of $B_1, B_3, \cdots, 1, 2$. 
($ Y_1, Y_2$ are written as $Y^{(1)}, Y^{(2)}$ in \cite{DTP2}.)

We are not starting from $Y$ but rather from the QTM.
Corresponding to breathers, we introduce ``breather" QTM by

\begin{eqnarray*}
T_{B_1} (x) &:=& T_1 (x)       
\\
T_{B_3} (x) &:=& \Lambda_{(8,1)}(x+\frac{13}{14}i)  /\phi(x-\frac{12}{7} i) 
\\
T_{B_5}(x) &:=& \Lambda_{(15,8,8)/(7,7)}(x)/\phi(x\pm \frac{3}{2} i) 
 \\
T_{B_7}(x) &:=& \Lambda_{(15,15,8,8)/(14,7,7)}(x+\frac{11}{14}i) /
 (\phi(x-\frac{12}{7} i)  \phi(x\pm \frac{9}{7}i))    
 \\
T_{B_2}(x) &:=& T_7 (x)  
\\
T^{(6)}(x) &:=& \Lambda_{(8,7)/(6)}(x+\frac{25}{14}i)  
\end{eqnarray*}
then the following relations, referred to as  the ``breather" $T$ system, hold.
\begin{eqnarray*}
T_{B_1} (x\pm \frac{3}{14} i)&=& T_0(x\pm \frac{11}{14}i) + \phi(x-\frac{12}{7} i) T_{B_3} (x)    \\
T_{B_3} (x\pm \frac{3}{14} i)&=&  T_0(x) T_0(x \pm \frac{8}{14}i)  + T_{B_1}(x) T_{B_5} (x)\\
T_{B_5} (x\pm \frac{3}{14} i)&=&  T_0(x\pm \frac{3}{14}i)  T_0(x\pm \frac{5}{14}i) +
   T_{B_3} (x) T_{B_7} (x)\\
T_{B_2} (x\pm \frac{3}{14} i)&=& T_0(x\pm \frac{1}{14}i)+ T^{(6)}(x) 
\end{eqnarray*}

where $T_0(x) = f_2(x)$.
They are originated from  the ``hidden su(2)"  discussed in \cite{ZPG}.

In  contrast to the  dilute $A_3$ model (equivalently the $E_8$ case),  the ``hidden su(2)"  
structure is not enough to obtain a closed  set of functional relations.
We then introduce another set of functional relations, related to magnons.

\begin{figure}[hbtp]
\centering
\includegraphics[width=4cm]{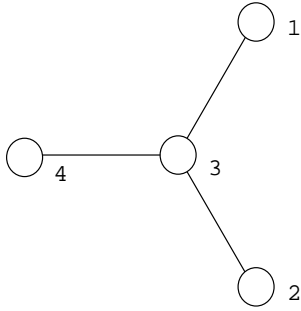}
\caption{ 
The nodes in the  $D_4$ Dynkin diagram are indexed in the above manner.
  }
\label {D4dynkin}
\end{figure}

 To each nodes on the $D_4$ Dynkin diagram (see fig \ref{D4dynkin}),  we associate
$t^{(a)}_m(x), (a=1,2,3,4, m \in Z_{\ge 0})$ and  set $t^{(a)}_0(x)=1$.
Then we impose a $D_4$ related $T$ system among them, in terminology of \cite{KNS1},
\begin{equation}
t^{(a)}_m (x \pm \frac{i}{14})= t^{(a)}_{m-1}(x) t^{(a)}_{m+1} (x) +g^{(a)}_m(x) \prod_{b \sim a}  t^{(b)}_m(x)
\label{D4like}
\end{equation}
where  $g^{(a)}_1 (x \pm \frac{i}{14})= g^{(a)}_2(x)$.   
In the above by $b \sim a$, we mean that $a$ and $b$ are connected on the  Dynkin diagram.

Moreover we set an inhomogeneous truncation, 
$t^{(3)}_3= t^{(4)}_3=0$ and put 
 $g^{(3)}_1=g^{(4)}_1=1$.

Unless one introduces some further condition, the set of functional relations (\ref{D4like})
are not closed, so can not be solved.
Then we demand 

\begin{eqnarray*}
t^{(1)}_3(x)&=& t^{(3)}_2(x),   \quad 
t^{(2)}_3(x)=  t^{(3)}_2(x) T_{B_3}(x) ,   \\
g^{(1)}_1(x) &=& T_0(x) , \quad g^{(2)}_1(x)=T_0(x\pm \frac{2 i}{7}) .
%
\end{eqnarray*}

The second relation  glues the breather $T$ system to the $D_4$ related $T$ system system.

The above requirements seem to be rather artificial, but they lead to remarkable consequences.
First,  solutions to  (\ref{D4like}) can be given in terms of QTM appearing in the dilute $A_5$ model
as follows.

\begin{alignat*}{2}
t^{(1)}_1(x) &= T^{(6)}(x)    
& \quad 
t^{(2)}_1(x) &= T_{B_7}(x)    
\\ 
t^{(3)}_1(x) &= \Lambda_{(12,8,7)/(5,4)}(x) 
& \quad  
t^{(4)}_1(x) &= \Lambda_{(5,1)}(x+\frac{15}{14}i)  /\phi(x-\frac{13}{2}i+\frac{15}{14}i) 
\\
t^{(1)}_2(x) &=T_{B_5}(x) T_{B_2}(x\pm \frac{i}{7})
& \quad
t^{(2)}_2(x) &= T_{B_5}(x \pm \frac{i}{7}) T_{B_2}(x)  
\\
t^{(3)}_2(x) &=T_{B_5}(x\pm \frac{i}{14}) T_{B_2}(x\pm \frac{i}{14})  
& \quad
t^{(4)}_2(x) &= T_{B_5}(x) T_{B_2}(x) . 
\end{alignat*}

The proof of the above statement is too lengthy to reproduce here.
We hope to present them with the general discussion of $L$ general \cite{DST}.

Second, the following combination of $T$ and $t$ solves the $Y-$ system for
$M_{5,6}+\phi_{1,2}$.

\begin{alignat*}{2}
Y_{B_1}(x)&=  \frac{ \phi(x-\frac{12}{7} i) T_{B_3} (x) }{  T_0(x\pm \frac{11}{14}i) }  
& \quad 
Y_{B_3}(x)&= \frac{T_{B_1} (x) T_{B_5} (x)}{T_0(x) T_0(x \pm \frac{8}{14}i) }   
\\
Y_{B_5}(x)&= \frac{T_{B_3} (x) T_{B_7}  }{ T_0(x\pm \frac{3}{14}i)  T_0(x\pm \frac{5}{14}i)}
& \quad 
Y_{B_2}(x)&=  \frac{T^{(6)} (x) }{T_0(x\pm \frac{1}{14}i) } 
 \\
Y_{K_1} (x)&= \frac{t^{(1)}_2(x)}{t^{(3)}_1(x) g^{(1)}_1(x)}   
& \quad 
Y_{K_2} (x)&= \frac{t^{(2)}_2(x)}{t^{(3)}_1(x) g^{(2)}_1(x)}  
\\
 Y_1 (x)&= \frac{t^{(3)}_2(x)}{t^{(1)}_1(x) t^{(2)}_1(x) t^{(4)}_1(x)}   
& \quad 
Y_2 (x)&= \frac{t^{(4)}_2(x)}{t^{(3)}_1(x) } .   
\end{alignat*}

Third, the functions $T$ , $t$, $Y$ possess ``nice" analytic properties. 
Before stating the properties, we need preparations.
Note that the $Y-$ system is invariant, for even $N$,  if
 $Y$ is replaced by $\widetilde{Y}$,  defined by
\begin{equation*}
\widetilde {Y}_{B_1}(x) =
  \begin{cases}
    \frac{Y_{B_1}(x) }{\kappa(x \pm i(1+u') \frac{3}{14})}&  \hbox{ for }   u<0 \\
   Y_{B_1}(x)  \kappa(x \pm i(1-u') \frac{3}{14})   &  \hbox{ for }   u>0 \\
  \end{cases}
\end{equation*}
and all other cases,   $\widetilde{Y_a}=Y_a$.  The parameter $u'$ stands for $\frac{14}{3} u$.
This is due to the definition of $\kappa$,
$$
\kappa(x) =\Bigl (
 i \frac{ \vartheta_1 ( i \frac{7}{6} \pi x, \tau') }
           {   \vartheta_2 ( i \frac{7}{6} \pi x, \tau') }
  \Bigr )^N
$$
which satisfies  $ \kappa(x\pm i \frac{3}{14})=1$.
The elliptic nome $q'=\exp(-\tau'), \tau'=4 \tau$ is introduced so as to 
respect the periodicity of the $Y$ function on the real direction of $x$. 
We denote  a typical  $\widetilde{Y}$ equation  as 
\begin{equation}
\widetilde{ Y_a} (x\pm i \alpha) = \prod_{b } \Xi_b (x \pm i  \gamma_b)   
           \prod_{c } {\cal L}_c(x \pm i  \gamma_c)     \label{typical}
\end{equation}

Our numerical data indicate that the rhs is analytic and nonzero in the strip $\Im x \in [-\alpha, \alpha]$.
Each element in the lhs also satisfies the same in appropriate strips , i.e.,   $\Xi_b(x)$ is 
analytic and nonzero in the strip $\Im x \in [-\gamma_b, \gamma_b]$, and so on.
These remarkable properties enable us to solve the coupled algebraic equation, like   (\ref{typical}), in the Fourier space
( to be precise, its logarithmic derivatives).
Then the inverse Fourier transformation leads to the coupled 
integral equations which yield the explicit evaluation of  $\log Y_a (x)$.

To make a direct contact with the TBA result, three further steps are needed.
First take the Trotter limit $N \rightarrow \infty, u N = \beta, (\epsilon=-1)$. 
Second rewrite  $\log \Xi_b (x) $ by  $\log {\cal L}_b(x) $.  Third, take a scaling limit.
The step1 is executable analytically, which manifests one of the advantage
of the present approach. 
The resultant equations no longer have dependency on a fictitious $N$ but only
depends on the temperature variable, $\beta$.
After the step 2,  we obtain the equations, in the Fourier space, 
$$
\widehat{M }
\begin{pmatrix}   
                        \widehat{\log} Y_{B_1} \\
                            \widehat{\log} Y_{B_3} \\
                           \vdots \\
\end{pmatrix} 
=4 \pi \beta 
 \begin{pmatrix}   1 \\
                           0 \\
                           \vdots \\
\end{pmatrix} 
+
\widehat{K_0 }
 \begin{pmatrix} 
                             \widehat{{\cal L}}_{B_1}  \\
                            \widehat{{\cal L}}_{B_3}   \\
                           \vdots \\
\end{pmatrix} 
$$
where $\widehat{{\cal L}}_{B_1} = \widehat{\log}(1+ \frac{1}{Y_{B_1}})$ and similarly for others.
The quantities with hat indicate that they are Fourier transformations.
 $\widehat{M }$ and $\widehat{K_0 }$ are asymmetric matrices of which explicit forms are omitted here
 but   can be easily obtained from the $Y$system.
  The only first entry has a nonvanishing inhomogeneous term in the rhs.
  This   reflects the fact 
 that  only $Y_{B_1}$ needs some trivial renormalization so as to have nice analytic properties.
By multiplying $M^{-1}$ from the left,  the kernel matrix of  TBA, 
 $M^{-1} K_0$  turns out to be symmetric, remarkably.
This property is crucial in applying the dilogarithm technique to evaluate the central charge.
The inhomogeneous term vector  $4\pi \beta M^{-1} \cdot  \phantom{  }^t (1,0,\cdots)$ possesses six non vanishing elements.

\begin{alignat*}{2}
\widehat{ d}_{B_1}&=\frac{ 8\pi \beta \cosh\frac{11}{14}k }{(2\cosh\frac{2}{14}k -1)D(k)}  
 \quad &
 \widehat{ d}_{B_3}&=\frac{ 4 \pi \beta (2 \cosh\frac{2}{14}k +1)  (2\cosh\frac{4}{14}k -1)   }{D(k)}    
 \\
 \widehat{ d}_{B_5}&=\frac{ 16 \pi \beta \cosh\frac{1}{14}k \cosh\frac{4}{14}k  }{(2\cosh\frac{2}{14}k -1) D(k)}  
 \quad  &
 \widehat{ d}_{B_2}&=\frac{ 8\pi \beta \cosh\frac{1}{14}k  }{(2\cosh\frac{2}{14}k -1) D(k)} 
 \\
 \widehat{ d}_{K_1}&=\frac{ 4\pi \beta }{(2\cosh\frac{2}{14}k -1) D(k)}   
 \quad &
\widehat{ d}_{K_2}&=\frac{ 8\pi \beta \cosh\frac{4}{14}k }{(2\cosh\frac{2}{14}k -1) D(k)} ,
 \end{alignat*}
where we denote by $\widehat{d}_{B_1}$ for the drive term associated to $\widehat{\log} Y_{B_1}$ and so on.
A common denominator $D(k)$ denotes
$$
D(k)=2 \cosh \frac{12}{14}k+2  \cosh \frac{10}{14}k -2  \cosh \frac{6}{14}k-2  \cosh\frac {4}{14}k+1.
$$

We finally perform the step 3. 
 In view of QFT, the  bulk  quantity is not of direct interest, rather
the fluctuation is. 
  We  introduce $y_{B_1}(x) = Y_{B_1}(x+\tau")$, for example, to  evaluate quantities near the
``fermi surface" with   $\tau"=\frac{12 \tau}{7 \pi}$.
  Then take a limit $q \rightarrow 0$  such that  $m_k R =\frac{8\pi \beta r}{2 \cos \frac{\pi}{21} -1} q^{\frac{4}{7}}$.
By $r$ we mean the residue of $i/D(k)$ at $k=\pi/3i$.
Two quantities $M^{-1}$ and $ K_0$ seem to carry the information of $S$ matrices;
the elements of   $M^{-1} K_0$   agree with the expression  described in \cite{DTP2} in terms of $S$ matrices,
under identification $x =3 \theta/\pi$  in the limit $q \rightarrow 0$.
The   matrix $M^{-1}$  also encodes the information of the mass spectra.
When taking the inverse Fourier transformation, the nearest zero to the real axis,
$k=\pm i\frac{\pi}{3}$ 
of $D(k)$, is   
relevant in the ``scaling" limit  as $ \tau"$ tends to be infinity.
Applying 
the Poisson's summation formula, we found a most dominant term, 
$$
d_{K_1}(x) = \frac{8\pi \beta r}{2 \cos \frac{\pi}{21} -1}  e^{-\frac{4}{7}\tau}  \cosh \frac{\pi}{3}x 
=m_K R \cosh \theta, 
$$
 for example, where   $\frac{\pi}{3}x  =\theta$.
Note that the relation $m_K \propto q^{\frac{4}{7}}$ is consistent with the scaling dimension
$\triangle_{1,2}=\frac{1}{8}$.
 One similarly verifies that all other drive terms also take the form $m R \cosh \theta$
and their mass ratio agree with those in \cite{DTP2}.
\begin{align*}
m_{B_1} &= 2 m_K \cos\frac{11}{42} \pi &   m_{B_3} &=4 m_K  \cos\frac{11}{42}\pi  \cos\frac{3}{42}\pi  \\
m_{B_5}&=4 m_K  \cos\frac{1}{42}\pi  \cos\frac{4}{42}\pi & 
m_{B_2} &= 2 m_K \cos\frac{1}{42} \pi   \\
m_{K_2} &= 2 m_K \cos\frac{4}{42} \pi     &  \\     
\end{align*}

Thus the TBA of  $M_{5,6}+\phi_{1,2}$ theory 
is recovered from  the scaling limit of the dilute $A_5$ model
at regime 2.

Once $Y$ is fixed by TBA,  we can also evaluate the free energy from
$$
T_1(x\pm \frac{3}{14}i)= T_{B_1}(x\pm \frac{3}{14}i) = T_0(x\pm \frac{11}{14}i) (1+Y_{B_1}(x)).
$$
It is readily shown that a ``fluctuation" part of the free energy $f$ is proportional to
$\frac{1}{\beta^2} \sum_k \int  m_k R \cosh \theta \log(1+1/y_k) d \theta$, which is the desired
expression.

\section{ dilute $A_2$ model at regime 1 as a lattice analogue to $M_{3,4}+\phi_{2,1}$ }

We treat another example corresponding to
$\phi_{2,1}$ perturbation theory,  the simplest and most well studied case, the Ising model
off critical temperature,  $M_{3,4}+\phi_{2,1}$. 
 The model is described by a free fermion, thus is rather
trivial in a sense. 
 In view of functional relations, however, it is not trivial to derive the simplest 
$Y$ system  $Y(x \pm i\frac{3}{2})=1$ (in the present normalization of $x$), from $T_1(x)$ in (\ref{dvf})
 which consists of 3 terms.
 This model is actually one of the first examples, which require a more fundamental object than $T_1(x)$, a box,
 which seems to correspond to a fundamental breather $B_1$.

 We define
 \begin{eqnarray}
\tau_K(x):&=&w \phi(x+2 i) \frac{Q(x+2i)}{Q(x+i)} +\phi(x) \frac{Q(x)Q(x+3i)}{Q(x+i)Q(x-i)} \\
&  & + w^{-1} \phi(x-2i) \frac{Q(x-2i)}{Q(x-i)}
\label{deftau}
\end{eqnarray}
which has a property common to $T_1(x)$ namely, it is 
pole-free due to the Bethe ansatz equation.   

More importantly, we have functional relations,
\begin{eqnarray}
\tau_K(x \pm \frac{1}{2}i)   &=& T_1(x)+T_2(x) = 2 T_1(x)  \label{IsingT1} \\
\tau_K(x \pm \frac{3}{2}i)  &=& T_3(x)+T_0(x)-   \phi(x\pm \frac{5}{2}i)(w^3 +\frac{1}{w^3})
= 2  (\phi(x\pm \frac{1}{2}i) + \phi(x\pm \frac{5}{2}i) )      \label{YY1}
\end{eqnarray}

The first equalities are directly verified by comparing both sides in forms of the ratio of $Q$ functions.
The seconds are consequences of the duality.  
One then reaches a desired relation  (\ref{YY1}) after proper renormalizations.
The first equation,  (\ref{IsingT1}) seems to suggest $\tau_K(x)$ is related to the kink in the theory;
the bound state of kink produces a breather.

In general $L=$ even case, we find  that $\tau_K(x)$ plays the most fundamental role, which will be a topic of 
 a separate publication. 

It is a nice exercise  to recover from (\ref{IsingT1}) and (\ref {YY1}), the free fermion free energy, in the scaling limit.
We shall remark the analytic property,  supported by numerics, that 
 $\tau_K(x)$ being Analytic  and Nonzero in the strip $Im x \in  [-\frac{3}{2}, \frac{3}{2}]$,
 for that purpose.

\section{Summary and discussion}

In this report, we demonstrate explicitly that  TBA  for $M_{5,6}+\phi_{1,2}$ and
 $M_{3,4}+\phi_{2,1}$, conjectured by Dorey et al, are realized in the scaling limit
 of lattice models.
The crucial idea is to introduce fusion transfer matrices associated to skew Young
tableaux and to investigate the functional relations among them.
The proofs of functional relations are rather combinatorial and lengthy, thus 
 omitted due to the lack of space.  They will be supplemented in the subsequent paper which 
 discusses the TBA behind the dilute $A_L$ models, $L$ general \cite{DST}.

There are still many open problems. 
The explicit identification of string solutions would be
definitely one of the most important.
The complete study on this will shed some lights on the 
way how to proceed  for TBA in the case of  perturbed non-unitary minimal
 models.
We mention the first step in this direction in \cite{EllemBaz2}.

 \vskip 0.9cm 
 \noindent The author would like to thank P. Dorey and R. Tateo for many valuable comments,
 discussions and collaborations,  S.O. Warnaar for useful correspondence.
  He also thanks for organizers of RAQIS03  for the
 nice conference and  kind hospitality.
 This work has been supported by a Grant-in-Aid for Scientific Research from the Ministry of
 Education, Culture, Sports and Technology of Japan, no. 14540376.


\clearpage
\begin{thebibliography}{10}

\bibitem{Zam891}
A.B. Zamolodchikov,  
Adv. Stud. Pure. Math. 19 (1989) 641.

\bibitem{Zam892}
A.B. Zamolodchikov, 
 Int. J. Mod. Phys. A 4 (1989) 4235.
 
 \bibitem{EguchiYang}
 T. Eguchi and S.K. Yang, Phys.Lett.B 235 (1990) 282-286.
 
 \bibitem{AlZamoRSOS}
 Al. B. Zamolodchikov, 
 Nucl. Phys. B358 (1991) 497-523.
 
 \bibitem {Smir}
 F. A. Smirnov, Int. J. Mod. Phys. A6 (1991) 1407-1428.
 
\bibitem {ChimZam} L. Chim and A. Zamolodchikov,
Int. J. Mod. Phys. A 7 (1992) 5317-5335.
 
 \bibitem{DTP1}
P. Dorey, A. Pocklington and  R. Tateo,
 Nucl. Phys. B{\bf 661} 425-463 (hep-th/0208111).


\bibitem{DTP2}
P. Dorey, A. Pocklington and  R. Tateo,
Nucl. Phys. B{\bf 661} 464-513  (hep-th/0208202).

\bibitem{Roberto95}
  R. Tateo,
Int. J.  Mod. Phys. A9 (1995) 1357-1376.


\bibitem{DTT}
P. Dorey, R. Tateo and  K.E. Thompson,
Nucl. Phys. B470  (1996) 317


\bibitem{WNS1}
S.O.Warnaar, B. Nienhuis and  K. A. Seaton,
Phys. Rev. Lett. 69(1992) 710.

\bibitem{WNS2} 
S.O.Warnaar, B. Nienhuis and  K. A. Seaton,
Int. J. Mod. Phys. B 7 (1993) 3727.

\bibitem{WPNS}
S.O.Warnaar, P.A. Pearce,  B. Nienhuis and  K. A. Seaton,
J. Stat. Phys 74 (1994) 469.

\bibitem{BS1} M.T. Batchelor and K.A. Seaton, 
J. Phys. A 30 (1997) L479.

\bibitem{KSeaton}  K.A. Seaton,  J.Phys. A35 (2002) 1597-1604

\bibitem{KorffSeaton}
C. Korff and K.A. Seaton,
Nucl.Phys. B636 (2002) 435-464




\bibitem{BWN}
V.V. Bazhanov, O. Warnaar and B. Nienhuis,
Phys. Lett. B 322 (1994) 198.

\bibitem{Suz85}
M.~Suzuki, Phys. Rev. B. 31 (1985) 2957.

\bibitem{Klu92}
A.~Kl\"umper, Ann. Physik 1 (1992) 540.


\bibitem{JSE8}
 J. Suzuki, 
Nucl Phys {\bf B528} (1998) 683.

\bibitem{JSE7}
 J. Suzuki, 
Progress in Math. 191 (2000) 217-247.


\bibitem{IK} A.G. Izergin and V. E. Korepin, 
Comm. Math. Phys. 79 (1981) 303.


\bibitem{Klu93}
See e.g., 
A.~Kl\"umper,
 Z. Phys. B 91 (1993) 507, 
 
 G.~J{\"u}ttner, A.~Kl{\"u}mper and J.~Suzuki, 
Nucl. Phys. B 512 (1998) 581.

A. Kuniba, K. Sakai and J. Suzuki,
Nucl. Phys. B 525 (1998) 597-626.

\bibitem{BR} V.V. Bazhanov and N. Yu Reshetikhin, 
J.Phys. A 23 (1990) 1477.

\bibitem{SuzG2}  J. Suzuki, Phys. Lett. A 195 (1994) 190.

\bibitem{KS}  A. Kuniba and J. Suzuki, 
Comm. Math. Phys.173 (1995) 225.


\bibitem{ZPG} Y.K. Zhou, P.A. Pearce  and  U. Grimm  Physica A 222 (1995) 261.

\bibitem{KNS1} A. Kuniba, T. Nakanishi and J. Suzuki, Int. J. Mod. Phys. A9 (1994) 5215-5266.

\bibitem{DST}
P. Dorey, J. Suzuki and R. Tateo, in preparation.


\bibitem{EllemBaz2}  R.M. Ellem and V.V. Bazhanov
Nucl.Phys. B647 (2002) 404-432




\end{thebibliography}
\end{document}